\documentclass{emulateapj}

\begin{document}

\title{\ion{C}{4} EMISSION-LINE DETECTION OF THE SUPERNOVA REMNANT RCW 114}

\author{I.-J. Kim\altaffilmark{1,2}}
\author{K.-W. Min\altaffilmark{1}}
\author{K.-I. Seon\altaffilmark{2}}
\author{W. Han\altaffilmark{2}}
\author{J. Edelstein\altaffilmark{3}}

\altaffiltext{1}{Korea Advanced Institute of Science and Technology,
305-701, Daejeon, South Korea; ijkim@kasi.re.kr}
\altaffiltext{2}{Korea Astronomy and Space Science Institute,
305-348, Daejeon, South Korea}
\altaffiltext{3}{Space Sciences Laboratory, University of California, Berkeley, CA 94720}

\begin{abstract}
We report the detection of the \ion{C}{4} $\lambda\lambda$1548, 1551 emission line in the region of the \object{RCW 114} nebula using the FIMS/SPEAR data. The observed \ion{C}{4} line intensity indicates that \object{RCW 114} is much closer to us than \object{WR 90}, a Wolf-Rayet star that was thought to be associated with \object{RCW 114} in some of the previous studies. We also found the existence of a small \ion{H}{1} bubble centered on \object{WR 90}, with a different local standard of rest velocity range from that of the large \ion{H}{1} bubble which was identified previously as related to \object{RCW 114}. These findings imply that the \object{RCW 114} nebula is likely an old supernova remnant which is not associated with \object{WR 90}. Additionally, the global morphologies of the \ion{C}{4}, H$\alpha$, and \ion{H}{1} emissions show that \object{RCW 114} has evolved in a non-uniform interstellar medium.
\end{abstract}


\section{INTRODUCTION}

\object{RCW 114} is a large H$\alpha$ filamentary nebula which is located at $(l, b) \sim (344\arcdeg, -5.5\arcdeg)$ with an angular diameter of $\sim$250$\arcmin$. Based on its optical appearance as well as the size and thickness of the filament, \citet{mea77} suggested that \object{RCW 114} is an old supernova remnant (SNR). From \ion{Na}{1} D2 absorption line analysis for seven B-type stars in the line-of-sight toward \object{RCW 114}, \citet{bed84} estimated the distance to this target to be less than 200 pc and suggested that it is a fossilized SNR which is coming into equilibrium with the surrounding interstellar medium (ISM). Using the H$\alpha$ profiles from four slit positions of the filamentary nebula, \citet{mea91} estimated the radial expansion speed to be 25--35 km s$^{-1}$ and argued that \object{RCW 114} is a remnant of \ion{Type}{2} SN in its momentum conserving phase, based on the kinetic energy of the shell and the far-infrared flux densities obtained from the {\it IRAS} data. A spectroscopic study of \object{RCW 114} confirmed that the optical emission is produced by the interaction of the shock wave of an SNR with the ambient ISM \citep{wal01}.

On the other hand, \citet{cap88} argued that \object{RCW 114} could be formed as a result of the strong stellar wind of the Wolf-Rayet (WR) star, \object{WR 90} (also known as HD 156385), and its massive progenitor, based on the newly discovered expanding \ion{H}{1} 21 cm bubble surrounding \object{RCW 114}. Recently, \citet{wel03} also proposed that \object{RCW 114} is associated with \object{WR 90}. Based on the observations of the interstellar \ion{Na}{1} D1 \& D2 absorption lines of seven early-type stars as well as the \ion{Si}{2} $\lambda$1304 absorption lines of four stars in the line-of-sight toward \object{RCW 114}, they argued that the absorption components are caused by the expanding stellar wind-blown bubble associated with \object{WR 90} while the optical emission arises in a shocked filamentary gas associated with the pre-existing evolved SNR cavity, which was probably formed by the high-mass progenitor of the present WR star. If the association between \object{RCW 114} and \object{WR 90} really exists, the distance to \object{RCW 114} should be similar to the distance to \object{WR 90} of $\sim$1.55 kpc \citep{des00}.

As soft X-ray emission has not been detected and there has been no estimation of the spectral index of the radio continuum, it is difficult to determine whether \object{RCW 114} is an SNR or not. However, even if \object{RCW 114} is an old SNR from which X-ray emission is not expected, we may still see some prominent far-ultraviolet (FUV) emission lines if its temperature remains at $\la$10$^{5}$ K, as the gas at this temperature is known to be inhabited by lithium-like C$^{+3}$, N$^{+4}$, and O$^{+5}$ ions, which have strong FUV resonance lines \citep{she98}. Especially, C$^{+3}$ ions can survive in a cooled gas of an old SNR longer than other high-stage ions \citep{she99} and their FUV resonance doublet is \ion{C}{4} $\lambda\lambda$1548, 1551. As can be seen in Table 2 of \citet{she98} and Figure 7 of \citet{she99}, \ion{C}{4} luminosity decreases much more slowly from its peak value than the X-ray luminosity as the SNR evolves. According to the simulation, 0.25 keV X-ray luminosity drops to $\sim$1\% of its peak value during a period of $\sim$2 $\times$ 10$^{6}$ yr; on the other hand, \ion{C}{4} luminosity remains at $\sim$10\% of its peak value until 1.2 $\times$ 10$^{7}$ yr. This means that \ion{C}{4} emission lines from $\la$10$^{5}$ K gas might be detectable for a considerable period in an old SNR that has already been cooled enough so that the soft X-ray from $\sim$10$^{6}$ K gas is too weak to be detectable. C$^{+3}$ ions can also be made by 48 eV photons from massive stars. However, the global morphology of the resulting \ion{C}{4} map can be considerably different from that of SNR origin.

In this paper, we present the detection of the \ion{C}{4} $\lambda\lambda$1548, 1551 emission line for the region of \object{RCW 114} using the data of the Far-Ultraviolet IMaging Spectrograph (FIMS), also known as Spectroscopy of Plasma Evolution from Astrophysical Radiation (SPEAR). We found an enhanced \ion{C}{4} region, which is closely associated with one of the H$\alpha$ filamentary features of \object{RCW 114}. As the FUV emission is highly extinguished by the interstellar dust, the observed \ion{C}{4} intensity could constrain the distance to \object{RCW 114}. In this way, we found that \object{RCW 114} is much closer to us than \object{WR 90}. We also compared the \ion{C}{4} and H$\alpha$ features with the \ion{H}{1} 21 cm morphologies using the \ion{H}{1} Southern Galactic Plane Survey (SGPS) data, which has a better resolution than that of the data used in \citet{cap88}. This study shows that \object{RCW 114} is not associated with \object{WR 90} and is likely an old SNR which has evolved non-uniformly.

\begin{figure}
\epsscale{1.1} \plotone{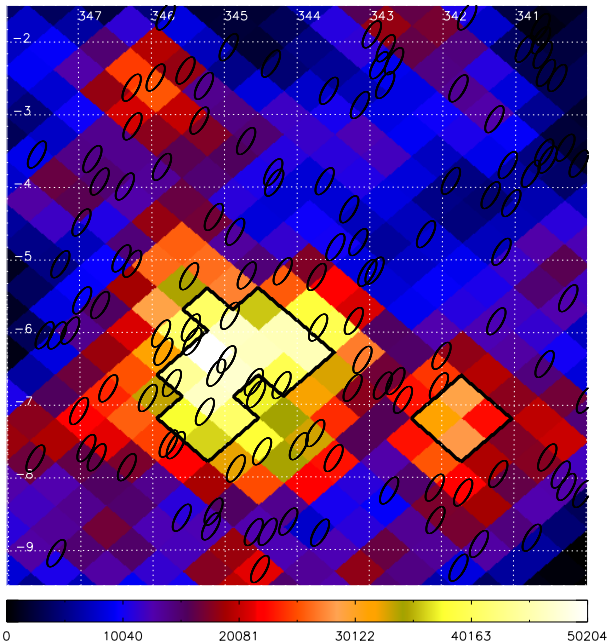} \caption{FIMS/SPEAR \ion{C}{4} $\lambda\lambda$ 1548, 1551 emission line image of RCW 114 in the Galactic coordinates. The unit of the color bar annotation is photons s$^{-1}$ cm$^{-2}$ sr$^{-1}$ (LU). The values are not corrected for interstellar extinction. The ellipses are the regions masked to remove stellar contamination. The contours indicate the signal-to-noise ratio of 3.0 for this image.\label{fig1}}
\end{figure}

\section{OBSERVATIONS AND DATA REDUCTION}

FIMS/SPEAR is the primary payload on the first Korean Science and Technology Satellite, {\it STSAT-1}, a micro-satellite launched on 2003 September 27. FIMS/SPEAR was designed to observe large-scale diffuse FUV emission lines from the ISM. FIMS/SPEAR consists of dual FUV imaging spectrographs: the short wavelength channel (S-channel; 900--1150 \AA, 4.0$\arcdeg$ $\times$ 4.6$\arcmin$ field of view) and the long wavelength channel (L-channel; 1340--1750 \AA, 7.4$\arcdeg$ $\times$ 4.3$\arcmin$ field of view), with $\lambda / \Delta \lambda \sim 550$ spectral resolution and 5$\arcmin$ angular resolution. The instrument, its on-orbit performance, and the basic processing of the data are described in detail in Edelstein et al. (2006a, 2000b).

We used the L-channel data of total 116 orbits obtained in the sky survey observational mode. Since our primary concern is the diffuse emission, we removed the data recorded when the count rate was high ($>$1000 counts s$^{-1}$), as these events were mostly associated with observation of bright stars. The photons of these bright stars were scattered and spread over the whole field of view, dominating any diffuse emission that might exist. We eventually obtained a total of $\sim$2.5 $\times$ 10$^{5}$ events with an average exposure time of $\sim$11.2 s for the area including \object{RCW 114}. The limited attitude accuracy of the satellite was augmented by automated software correction using the positions of the bright stars listed in the {\it TD-1} catalog \citep{tho78}. The positions of the reference stars were corrected to be accurate within 5$\arcmin$ in the present study. Then, the position errors of all photons were corrected by interpolating linearly from those of the bright stars. This method is similar to that used in \citet{seon06}. To obtain images and spectra, we have adopted the HEALPix scheme \citep{gor05} with the resolution parameter \texttt{Nside} = 2048, corresponding to a pixel size of $\sim$1.72$\arcmin$ and 3 \AA{} wavelength bins. The effects of bright stars were further reduced by masking the pixels of the elliptical region around the stars, with its major and minor axes of 25$\arcmin$ and 10$\arcmin$, respectively, based on the scattering pattern of the slit image. We identified a total of 125 stars in the area of present concern using the {\it TD-1} catalog and the {\it Tycho-2} spectral type catalog \citep{wri03}. The masked positions are indicated in Figure 1.

\begin{figure}
\epsscale{1.1} \plotone{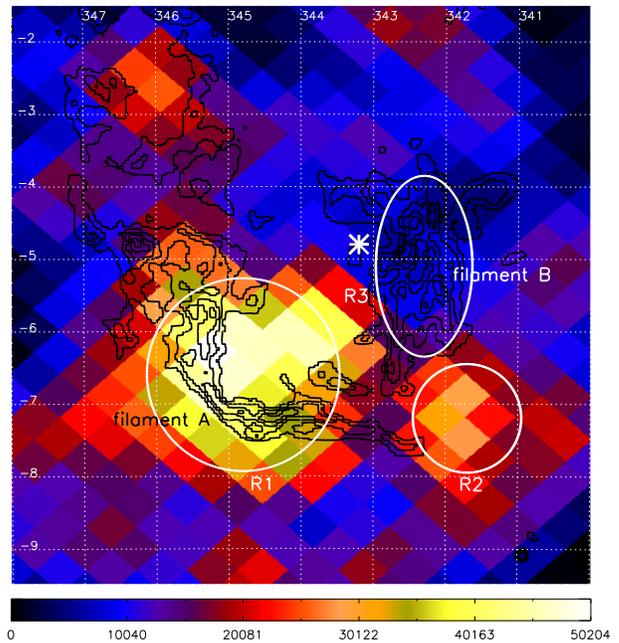} \caption{H$\alpha$ contours overlaid on the FIMS/SPEAR \ion{C}{4} $\lambda\lambda$ 1548, 1551 emission line image. The H$\alpha$ contours are from 70 to 190 rayleighs with 20 rayleigh intervals. The position of WR 90 is marked by an asterisk. The spectra in Fig. 3 are extracted from the three subregions designated by R1 through R3.\label{fig2}}
\end{figure}

\section{DATA ANALYSIS AND RESULTS}

For the \ion{C}{4} emission line image, we took the 1531--1570 \AA{} portion of the L-channel spectrum and fitted it with a constant continuum plus \ion{C}{4} $\lambda\lambda$1548, 1551 lines for each pixel. For better statistics, we increased the pixel size by 16 times (\texttt{Nside} = 128, corresponding to a pixel size of $\sim$27.5$\arcmin$). We assumed a 2:1 line ratio for 1548 and 1551 \AA{} doublet lines, as in an optically thin case \citep{seon06}. The model spectrum was convolved with a Gaussian function with the width of the FIMS/SPEAR spectral resolution. Because the signal from each pixel was not strong, the image was then smoothed using the spherical version of a Gaussian kernel with full width at a half-maximum of 60.0$\arcmin$ \citep{seo06} so that \ion{C}{4}-detected regions with $>$3 $\sigma$ confidence could be obtained. However, the \ion{C}{4}-detected regions should be double-checked because this smoothing method assumes Gaussian statistics and the result for the region with poor count statistics may not be reliable.

\begin{figure}
\epsscale{1.1} \plotone{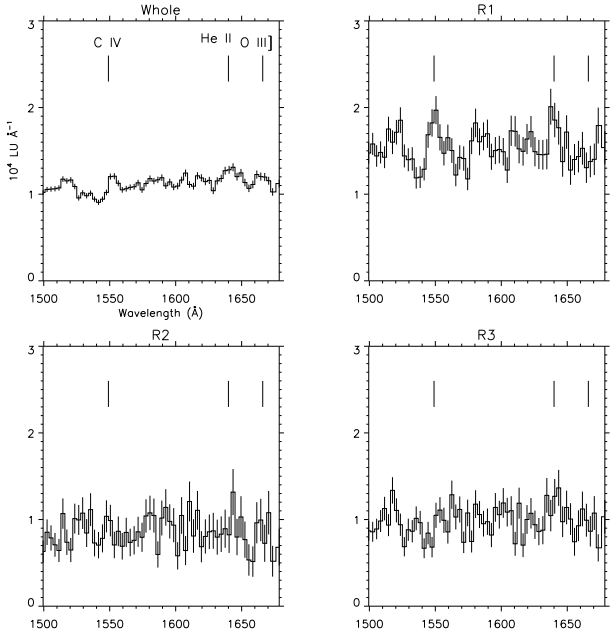} \caption{FIMS/SPEAR L-channel spectra (with 1 $\sigma$ error bars) from the whole region and the three subregions indicated in Fig. 2. The spectra are binned at 3 \AA{} intervals.\label{fig3}}
\end{figure}

Figure 1 shows the final result of the \ion{C}{4} emission line image. As can be seen in the figure, there is a strong \ion{C}{4} region at $(l, b) \sim (345.2\arcdeg, -6.3\arcdeg)$ with its peak value of $\sim$5.0 $\times$ 10$^{4}$ photons s$^{-1}$ cm$^{-2}$ sr$^{-1}$ (hereafter LU). Although the boundary of this strong \ion{C}{4} region is ambiguous due to smoothing, the emission patterns for this region in the original unsmoothed image are not dominated by only a few pixels and essentially the same as those in the smoothed image. The weaker \ion{C}{4} region at $(l, b) \sim (341.7\arcdeg, -7.3\arcdeg)$ is also identified with $>$3 $\sigma$ confidence in this smoothed image. However, we believe the detection from this weaker region is marginal as its intensity is low and the confidence level of the \ion{C}{4} emission line in the spectrum is $<$2 $\sigma$, as will be mentioned below. In Figure 2, H$\alpha$ contours in the range of 70--190 R are overplotted on the \ion{C}{4} emission line image. The H$\alpha$ contours show two filamentary features, the sharp one in the southeast (H$\alpha$ filament A) and the broad one in the west (H$\alpha$ filament B). The position of \object{WR 90} is near H$\alpha$ filament B, as marked by an asterisk in the figure. Interestingly, while the strong \ion{C}{4} region appears near the inside of H$\alpha$ filament A and is extended over to the filament, there is no enhanced \ion{C}{4} feature near H$\alpha$ filament B. We have marked three interesting regions with designations R1 through R3: the strongest \ion{C}{4} region including H$\alpha$ filament A (R1), the weaker \ion{C}{4} region (R2), and H$\alpha$ filament B (R3). The archival data of H$\alpha$ was adopted from the SkyView virtual observatory \citep{mcg98}.

In Figure 3, we have plotted the spectra for the whole region and subregions R1 through R3. The detector background was subtracted in these spectra despite their small size: $\sim$0.1\% of the lowest continuum level of the whole region or $\sim$7.3\% of the continuum level of region R2. The spectra were binned at 3 \AA{} intervals and shown with 1 $\sigma$ error bar in the figure. The spectrum for the whole region shows the \ion{C}{4} $\lambda\lambda$1548, 1551 doublet, which is, however, unresolved due to 3 \AA{} binning, in addition to rather weak ionic lines such as \ion{He}{2} $\lambda$ 1640, and \ion{O}{3}] $\lambda$ 1666. However, the \ion{He}{2} or \ion{O}{3}] line does not appear clear, and no spatial feature of either line appearing to be correlated with \object{RCW 114} was found. We fitted the spectrum for each subregion to obtain the \ion{C}{4} emission line intensity and the results are $(4.1 \pm 1.0) \times 10^4$ LU for R1, $(1.4 \pm 0.9) \times 10^4$ LU for R2, and $(0.0 \pm 0.3) \times 10^4$ LU for R3. For region R1, the \ion{C}{4} emission line intensity was determined with $>$4 $\sigma$ confidence while the confidence level is $<$2 $\sigma$ for region R2.

\begin{figure}
\epsscale{1.1} \plotone{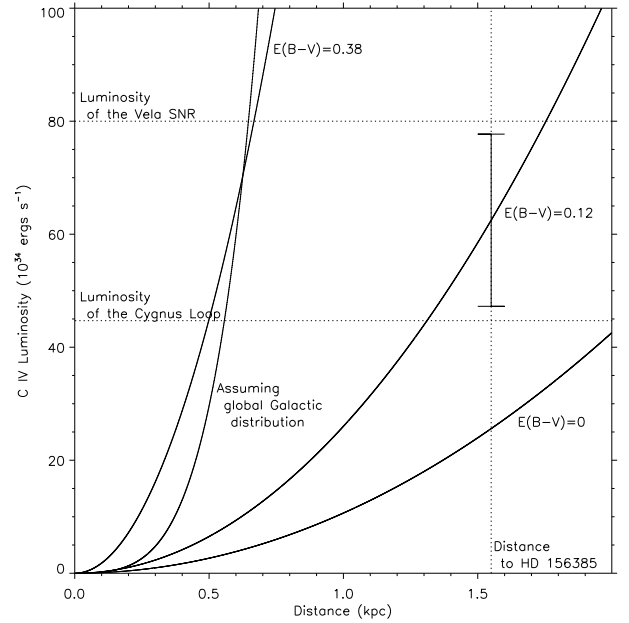} \caption{\ion{C}{4} line luminosities versus the assumed distance to RCW 114. The three curves represent the estimated C IV luminosities with reddening-correction by assuming the global Galactic distribution of dust, $E(\bv) = 0.12$, and $E(\bv) = 0.38$, respectively. A 1 $\sigma$ error bar is marked on the curve with $E(\bv) = 0.12$ at 1.55 kpc. The value for the distance to WR 90 is from \citet{des00}. The values for the \ion{C}{4} luminosities of the Vela SNR and the Cygnus Loop are from \citet{nis06} and \citet{seon06}, respectively.\label{fig4}}
\end{figure}

We estimated the \ion{C}{4} luminosity of region R1 for various assumed distances to the target, with or without reddening-corrections. For reddening-correction, we adopted three different methods to calculate the foreground extinction. Firstly, we assumed a global Galactic distribution model of dust and then estimated the foreground extinction from the column density of dust to region R1. Adopting the two-components model of \citet{mis06} and assuming the vicinity of the Sun (8 kpc from the Galactic center) that lies exactly on the Galactic plane, the column density of dust to a target which has the Galactic latitude $b$ and the distance $d$ is
\begin{eqnarray*}
\int_{0}^{d} [n_{\mathrm w, 0}\;e^{-(\frac{r\sin{|b|}}{z_{\mathrm w, 0}})-(\frac{8\ \mathrm{kpc}}{R_{\mathrm w, 0}})}\qquad\qquad\qquad\qquad\qquad \\ \qquad\qquad + n_{\mathrm c, 0}\;e^{-(\frac{r\sin{|b|}}{z_{\mathrm c, 0}})-(\frac{8\ \mathrm{kpc}}{R_{\mathrm c, 0}})}] \; dr \quad (\textrm{gr cm}^{-2}),
\end{eqnarray*}
where $n_{\mathrm w, 0} = 1.22 \times 10^{-27}$ gr cm$^{-3}$, $z_{\mathrm w, 0} = 0.09$ kpc, $R_{\mathrm w, 0} = 3.3$ kpc, $n_{\mathrm c, 0} = 1.51 \times 10^{-25}$ gr cm$^{-3}$, $z_{\mathrm c, 0} = 0.1$ kpc, and $R_{\mathrm c, 0} = 5.0$ kpc from Table 2 in \citet{mis06}. Multiplying the extinction cross section of 6.7 $\times$ 10$^{4}$ cm$^{2}$ gr$^{-1}$ for the Galactic dust with $R_V$ = 3.1 \citep{dra03} and substituting $b = -6.6\arcdeg$ for region R1, the foreground extinction can be represented as a function of the distance. Secondly, we simply adopted the observed value of $E(\bv)$ for \object{WR 90} (0.38; Dessart et al. 2000) and applied it regardless of the distances. Finally, we used the data in the catalog of extinctions and distances by \citet{gua92} to estimate the foreground extinction with various distances. A total of 18 stars are listed in the catalog for region R1 together with the distance information, except for five stars for which we used the distance data from the Hipparcos satellite (ESA 1997). The effective average value of $E(\bv)$ for various stars can be calculated from the effective optical depth, $\tau_{\mathrm{eff}} = -\ln<e^{-\tau_i}>$. We calculated the effective average values of $E(\bv)$ for 15 stars within 0.75 kpc, 16 stars within 1.0 kpc, and all 18 stars within 1.25 kpc, and obtained the same value of 0.12 for all three cases. Using these estimations of the foreground extinction and the extinction curve of \citet{car89} with $R_V$ = 3.1, three curves for the estimated \ion{C}{4} luminosities were plotted in Figure 4, along with the case without reddening correction. The implications of these curves, representing the estimated \ion{C}{4} luminosity as a function of distance, will be discussed in Section 5.

\begin{figure}
\epsscale{1.15} \plotone{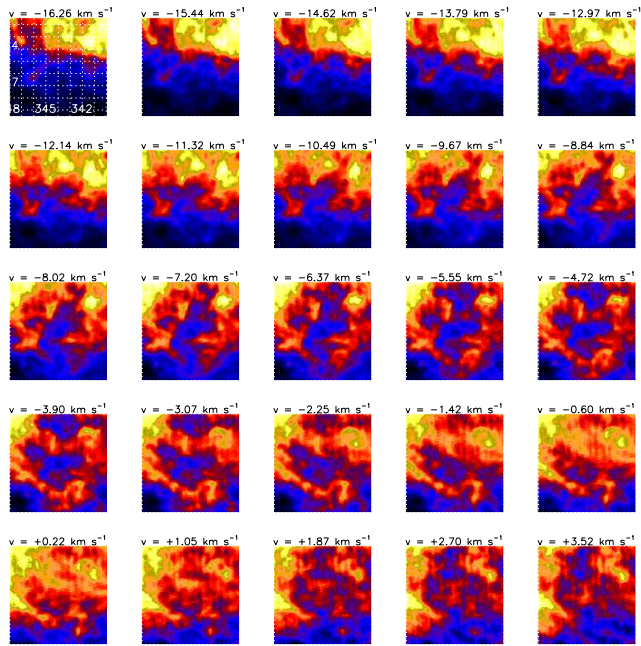} \caption{SGPS \ion{H}{1} 21 cm morphologies toward the RCW 114 region for LSR velocity range $-16.26 \leq v \leq +3.52$ km s$^{-1}$. The LSR velocity is indicated above each map. The projection type of each map is orthographic.\label{fig5}}
\end{figure}

\section{\ion{H}{1} 21 cm DATA}

To compare the present \ion{C}{4} emission line results with \ion{H}{1} 21 cm morphologies, we employed the Parkes survey data, which is part of the \ion{H}{1} Southern Galactic Plane Survey (SGPS; McClure-Griffiths et al. 2005), a survey of the \ion{H}{1} spectral line and 21 cm emission in the fourth quadrant of the Galactic plane. The Parkes data have an angular resolution of $\sim$15$\arcmin$ and a spectral resolution of 0.82 km s$^{-1}$, which are $\sim$3--4 times better than those of the data used in \citet{cap88}. FITS format data, consisting of orthographic projected images on the Galactic coordinates for each local standard of rest (LSR) velocity, are provided online (\url{http://www.atnf.csiro.au/research/HI/common/}).

\begin{figure}
\epsscale{0.55} \plotone{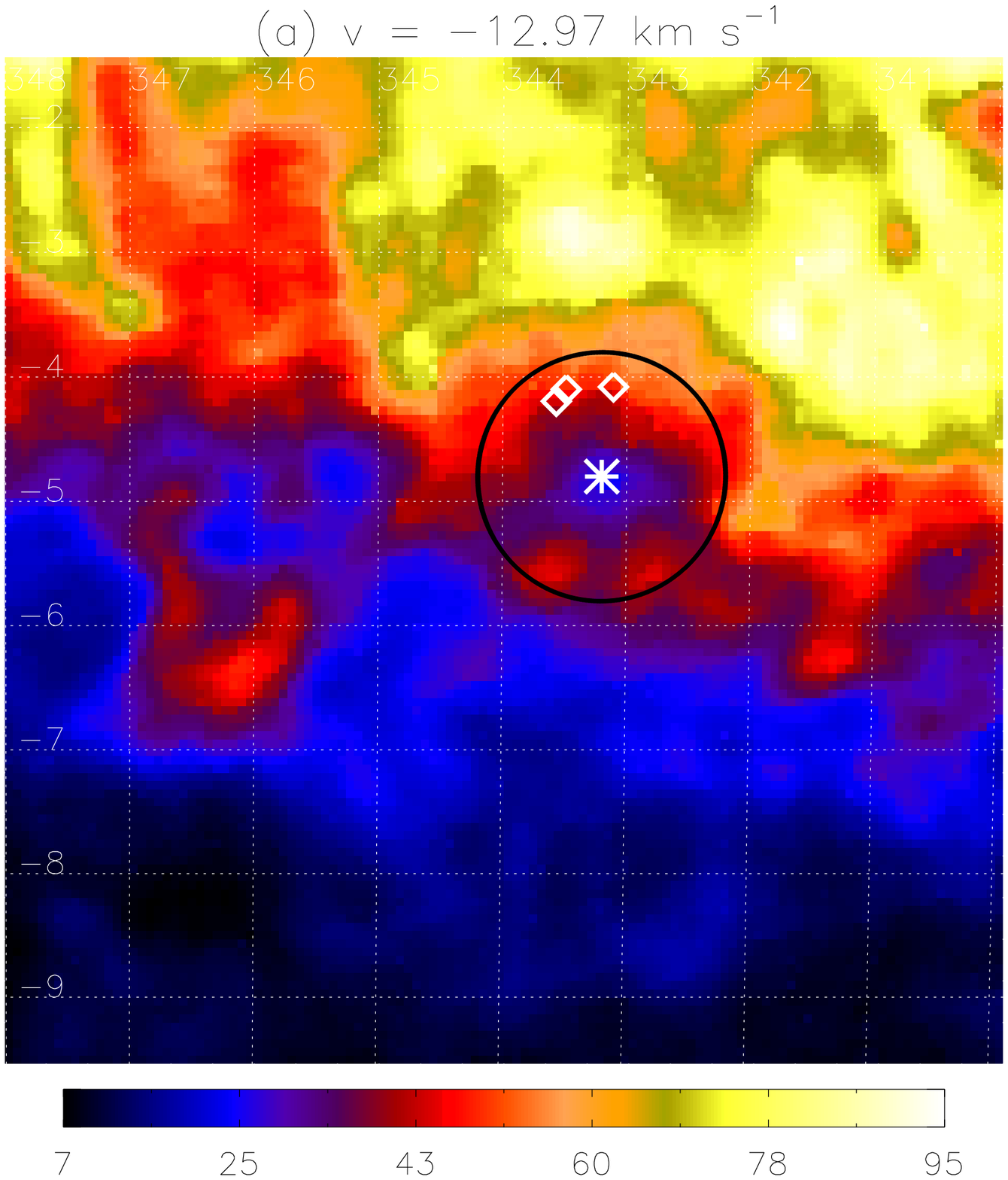} \plotone{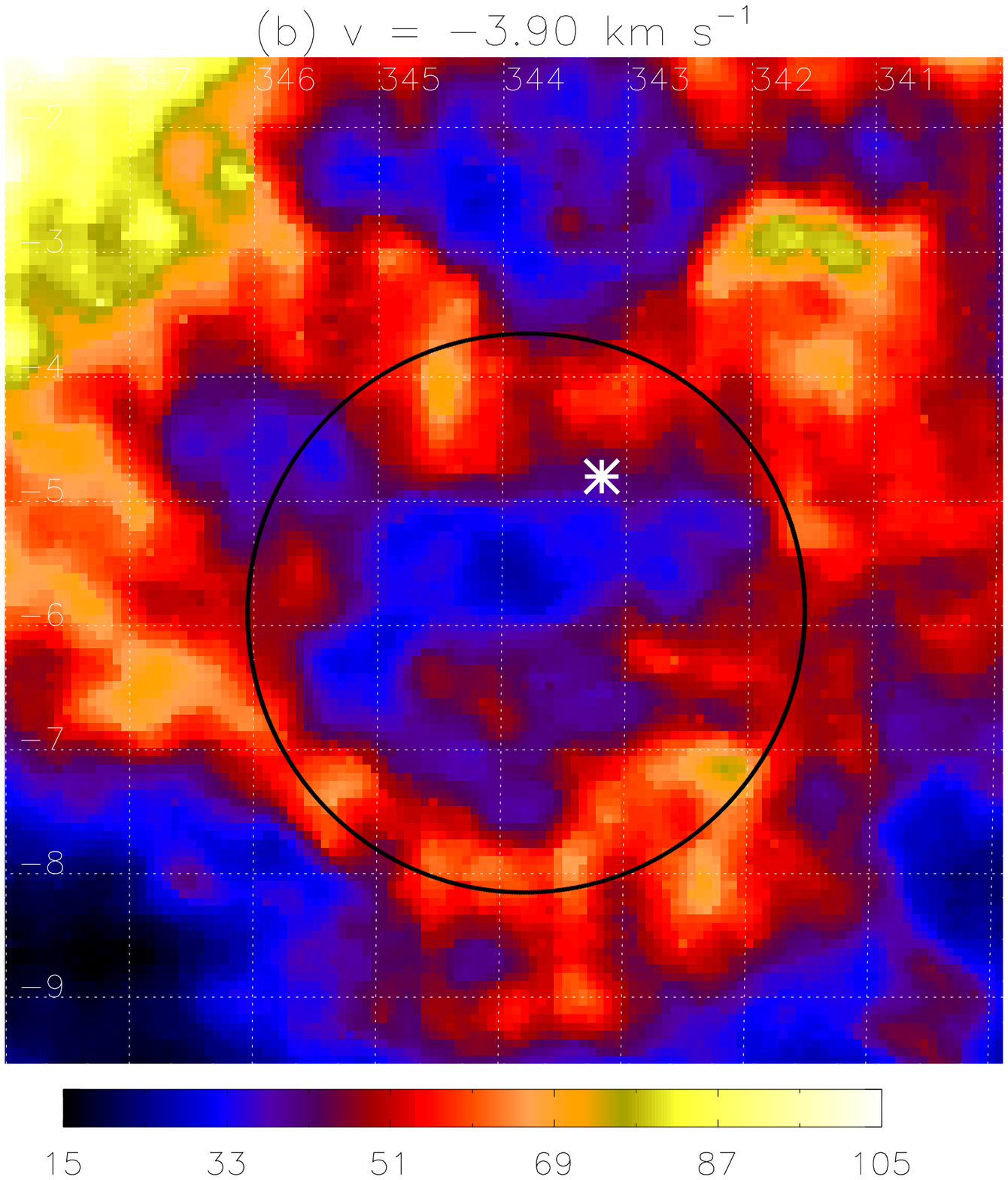} \plotone{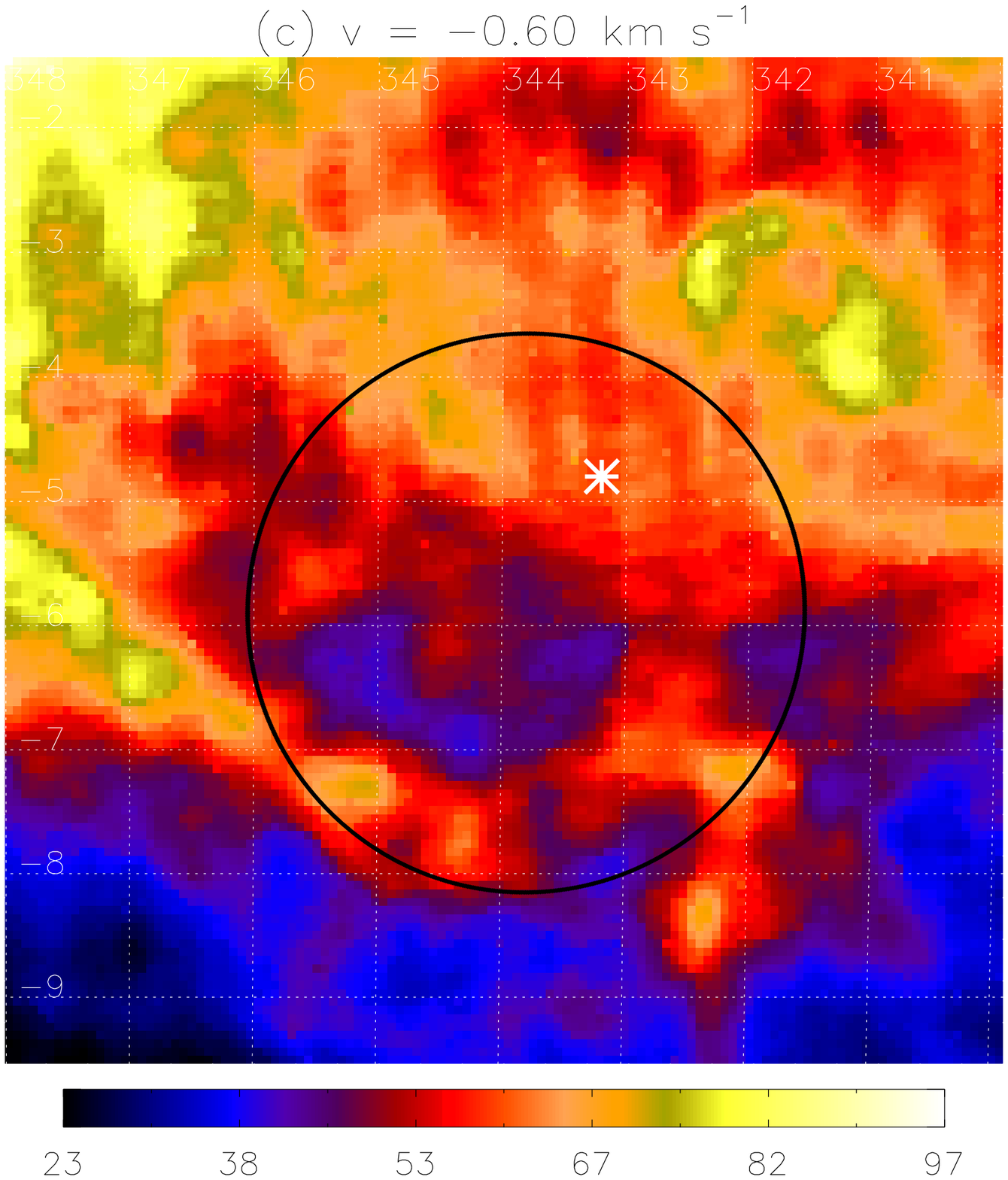} \plotone{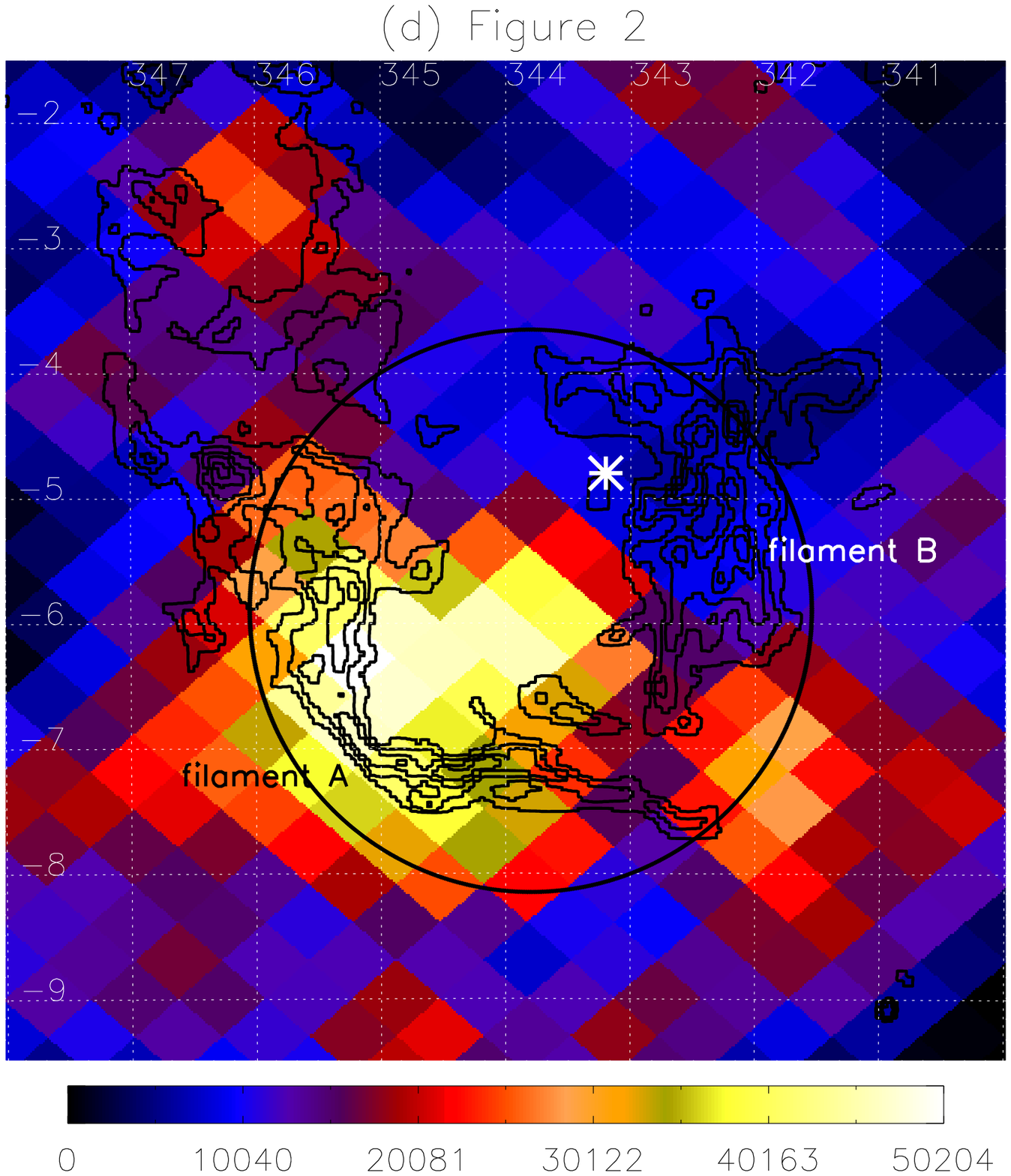} \caption{Comparison between ($a$)--($c$) SGPS 21 cm images and ($d$) Fig. 2. The corresponding LSR velocities are indicated above ($a$)--($c$) images. The unit of each color bar annotation is ($a$)--($c$) K and ($d$) LU. The projection type of each image is ($a$)--($c$) orthographic and ($d$) gnomic. The position of WR 90 is marked by an asterisk in each image. A circle with a radius of 1$\arcdeg$ in ($a$) is centered on this WR star, and each circle with a radius of 2.25$\arcdeg$ in ($b$)--($d$) is centered at $(l, b) \sim (343.8\arcdeg, -5.9\arcdeg)$. The three positions mentioned in \citet{wal01}, which have the [\ion{S}{2}]/H$\alpha$ line ratios typical of \ion{H}{2} regions, are indicated by diamonds in ($a$).\label{fig6}}
\end{figure}

Figure 5 shows \ion{H}{1} 21 cm morphologies toward the \object{RCW 114} region for LSR velocity range $-16.26 \leq v \leq +3.52$ km s$^{-1}$. While we used a different projection type for the maps in Figure 5 from that of Figure 2, the projected areas are more or less the same. Even a cursory examination of Figure 5 reveals that the morphology varies greatly as the LSR velocity changes. To emphasize the features associated with \object{RCW 114} or \object{WR 90}, we selected three velocity ranges of Figure 5 and compared the resulting \ion{H}{1} images with the \ion{C}{4} image of Figure 2 in Figure 6. In Figure 6$a$, we have drawn a circle with a radius of 1$\arcdeg$ centered on \object{WR 90} marked by an asterisk, and this coincides well with an \ion{H}{1} void at $v = -12.97$ km s$^{-1}$. The void looks especially clear on the northern limb of the circle for velocities ranging from $-14.62$ to $-11.32$ km s$^{-1}$ in Figure 5. Another \ion{H}{1} void indicated by a circle with a radius of 2.25$\arcdeg$ in Figures 6$b$--6$d$ shows the best agreement with the H$\alpha$ filamentary features at $v = -3.90$ km s$^{-1}$, as can be seen in Figures 6$b$ and 6$d$, although the associated features are noticeable at wider velocity ranges (from about $-8.02$ to about +1.05 km s$^{-1}$) in Figure 5. As the velocity increases from $v = -3.90$ km s$^{-1}$ in Figure 5, the void progressively shrinks in the southeastern part, and it becomes more or less coincident with the enhanced \ion{C}{4} region (R1), as can be seen in Figures 6$c$ and 6$d$.

\begin{figure}
\epsscale{1.1} \plotone{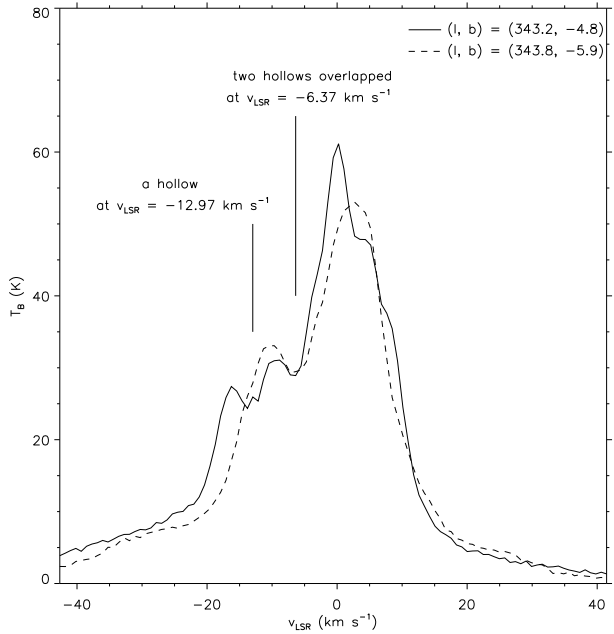} \caption{Velocity profiles for the centers of the two \ion{H}{1} bubbles: one for the position of the Wolf-Rayet star marked in Fig. 6$a$ ({\em solid line}) and the other for the center of the circle indicated in Fig. 6$b$ ({\em dashed line}). The two positions of the identified hollows are also indicated. \label{fig7}}
\end{figure}

To confirm the existence of the two independent \ion{H}{1} bubbles, we made LSR velocity profiles for the centers of the two assumed \ion{H}{1} bubbles: one for the position of \object{WR 90} marked in Figure 6$a$ and the other for the center of the circle indicated in Figure 6$b$. As the results are compared in Figure 7, both profiles have hollows overlapped at the same LSR velocity of $-6.37$ km s$^{-1}$, corresponding to the large \ion{H}{1} bubble indicated in Figure 6$b$. On the other hand, the velocity profile for \object{WR 90} indeed shows an additional hollow at $v = -12.97$ km s$^{-1}$, which confirms the existence of the smaller \ion{H}{1} bubble indicated in Figure 6$a$.

\section{DISCUSSION}

We have found an enhanced \ion{C}{4} region (R1) which seems to be correlated with H$\alpha$ filament A although it looks more extensive due to smoothing, as can be seen in Figure 2. Moreover, the \ion{H}{1} void of $v = -3.90$ km s$^{-1}$ correlated clearly with the H$\alpha$ filaments in Figures 6$b$ and 6$d$, shrinks at higher LSR velocities to become coincident with this enhanced \ion{C}{4} region, as can be seen in Figures 6$c$ and 6$d$. These correlations between the features seen in three different wavelength domains strongly suggest that they are physically associated with each other.

The \ion{C}{4} line luminosity estimated in the section above may constrain the possible origin of the \object{RCW 114} nebulosity, particularly in connection with \object{WR 90}. The \ion{C}{4} line luminosity from an old SNR or from an \ion{H}{2} region should be much lower than that of the \object{Cygnus Loop}, which is a well-known middle-aged SNR with a strong X-ray emission. As can be seen in Figure 4, the estimated \ion{C}{4} luminosity with reddening-correction by assuming the global Galactic distribution of dust, exceeds that of the \object{Cygnus Loop} (44.7 $\times$ 10$^{34}$ ergs s$^{-1}$; Seon et al. 2006) for a distance of more than $\sim$0.5 kpc. This means that the distance to \object{RCW 114} is much closer to us than \object{WR 90} whose distance is $\sim$1.55 kpc \citep{des00} unless the foreground extinction along the line of sight toward region R1 is much lower than expected from the global Galactic distribution of dust. When $E(\bv)$ = 0.38 for \object{WR 90} is adopted, the estimated \ion{C}{4} luminosity still exceeds that of the \object{Cygnus Loop} for a distance of more than $\sim$0.5 kpc and becomes $\sim$10 times larger than that of the \object{Cygnus Loop} at 1.55 kpc. The foreground extinction toward region R1 could possibly be lower than toward \object{WR 90}. If we, therefore, adopt $E(\bv)$ = 0.12 obtained from the 18 stars within region R1, the estimated \ion{C}{4} luminosity with the assumed distance of 1.55 kpc becomes similar to that of the \object{Cygnus Loop} at its lower limit of the 1 $\sigma$ error, which is related to the error in the estimation of \ion{C}{4} line intensity for region R1. However, as those stars within the specified distances (0.75, 1.0, and 1.25 kpc) were employed for the estimation of $E(\bv)$, the above value of $E(\bv)$ = 0.12 is actually a lower limit of the foreground extinction along the line of sight toward the target located behind the corresponding distances. Therefore, the \ion{C}{4} luminosity reddening-corrected by $E(\bv)$ = 0.12 is also a lower limit of the luminosity for the distance of 1.55 kpc. Hence, if the \ion{C}{4} emission line of region R1 comes from the same distance ($\sim$1.55 kpc) as \object{WR 90}, its total luminosity more likely exceeds that of the \object{Cygnus Loop}. In conclusion, the association of \object{RCW 114} with \object{WR 90} or its progenitor seems implausible and \object{RCW 114} is likely to be much closer to us than the star.

Furthermore, no direct association between \object{RCW 114} and \object{WR 90} is evidenced by the results of the \ion{H}{1} 21 cm observations. As mentioned in Section 4, we found another small \ion{H}{1} void toward \object{WR 90}. We believe the newly identified \ion{H}{1} void centering on \object{WR 90} in Figure 6$a$ and the previously identified \ion{H}{1} void \citep{cap88} in Figure 6$b$ are the two independent bubbles associated with \object{WR 90} and \object{RCW 114}, respectively. The radius of the smaller \ion{H}{1} void ($\sim$1$\arcdeg$) is equivalent to the radius of $\sim$27 pc at a distance of 1.55 kpc, which is compatible with those of the \ion{H}{1} bubbles found around other WC-type WR stars \citep{nie91}. We also note that the systemic LSR velocity of the smaller \ion{H}{1} bubble associated with \object{WR 90} is faster than that of the larger \ion{H}{1} bubble associated with \object{RCW 114}. The smaller \ion{H}{1} void appears largest at $v \sim -12.97$ km s$^{-1}$ in Figure 5 and the velocity profile passing through the center of this void shows a hollow exactly at the same velocity in Figure 7, meaning that $-12.97$ km s$^{-1}$ is the systemic velocity of the expanding \ion{H}{1} bubble. On the other hand, the systemic velocity of the larger \ion{H}{1} bubble is around $-6.37$ km s$^{-1}$, where the hollows of the two velocity profiles coincided in Figure 7, although the \ion{H}{1} void shows the best agreement with the H$\alpha$ filaments at $v \sim -3.90$ km s$^{-1}$. In view of the Galactic rotation, the approaching speed increases with distance within a few kpc in the direction of $l \sim 344\arcdeg$ on the Galactic plane. Hence, this is again consistent with our argument based on the \ion{C}{4} emission study that \object{RCW 114} is closer than \object{WR 90}.

When the [\ion{S}{2}] $\lambda\lambda$6717, 6731/H$\alpha$ line ratio is greater than 0.5, it is believed that the emission is produced by shock-excited gas \citep{fes85}. On the other hand, the ratio is usually much smaller in \ion{H}{2} regions and planetary nebulae. The three positions mentioned in \citet{wal01}, which have the [\ion{S}{2}]/H$\alpha$ line ratios typical of \ion{H}{2} regions, precisely fit the northern envelope of the smaller \ion{H}{1} void, as denoted by three diamonds in Figure 6$a$. This implies that the apparent northern envelope of the \ion{H}{1} bubble is photoionized by \object{WR 90}. The broadness of H$\alpha$ filament B could be due to the overlap by the \ion{H}{2} region around \object{WR 90} behind \object{RCW 114}. Two stars with a complex of \ion{Na}{1} D1 \& D2 absorption lines in \citet{wel03} are also located within the smaller \ion{H}{1} void of Figure 6$a$. Therefore, these complexities may not arise from their distances, as suggested, but may rather result from the positions of these stars which happen to be along the line of sight towards the \ion{H}{1} bubble created by \object{WR 90}. One of these stars is just \object{WR 90}.

One more interesting feature is that strong \ion{C}{4} emission appears near H$\alpha$ filament A while no enhanced \ion{C}{4} feature is seen near H$\alpha$ filament B in Figure 2. Hence, if the extinctions towards these two regions are not too different, the difference in the \ion{C}{4} emission should be the result of the non-uniform evolution of \object{RCW 114}. In fact, \citet{wal01} found that the [\ion{S}{2}]/H$\alpha$ line ratio, which is more or less independent of extinction, appears to be larger along H$\alpha$ filament A than along H$\alpha$ filament B and that it is in the range 0.5--0.7 for about half of the observed positions. This result led them to suggest that the expanding shock is interacting more strongly with the ISM in the region of H$\alpha$ filament A away from the Galactic plane. Though the boundary of the enhanced \ion{C}{4} region (R1) is rather ambiguous because of smoothing, this region seems to be more prominent inside the filament than outside. This broad \ion{C}{4} region could be a zone of cooling and recombining C$^{+3}$ ions which is expected to be developed widely in a fairly cooled SNR \citep{she98}. Therefore, we believe \object{RCW 114} is an old SNR which has evolved in a non-uniform ambient ISM with a lower-density southeast region. This SNR seems to have been cooled enough now so that soft X-ray emission is undetectable and \ion{C}{4} emission is seen only in the south and east regions, where the gas of $\la$10$^{5}$ K may still remain because the ambient ISM of lower density may have allowed the SNR to evolve slowly there.

The morphology of the \ion{H}{1} void associated with \object{RCW 114} can support the above scenario. A long and thin \ion{H}{1} feature seen at LSR velocities ranging from $-4.72$ to $-0.60$ km s$^{-1}$ appears outside H$\alpha$ filament A and the enhanced \ion{C}{4} region, as can be seen in Figures 6$b$--6$d$. If this \ion{H}{1} filament were a cool shell of the \object{RCW 114} SNR, it should be inside H$\alpha$ filament A because cool shells form behind evolved shocks of SNRs \citep{she98}. Therefore, this suggests that the \ion{H}{1} filament is not the cool shell of \object{RCW 114} and the whole \ion{H}{1} cavity containing \object{RCW 114} has likely pre-existed the SN explosion which led to the \object{RCW 114} SNR. As many cases of other SNRs, \object{RCW 114} may have evolved in this pre-existing cavity. We also note that the \ion{H}{1} void feature remains in the southeast part as LSR velocity increases from the systemic (central) LSR velocity of $-6.37$ km s$^{-1}$ to higher values, as mentioned in Section 4. Considering the radial structure of expanding \ion{H}{1} bubble centered at $v \sim -6.37$ km s$^{-1}$, the H I features at these higher velocities ($v > -6.37$ km s$^{-1}$) correspond to the rear side of the \ion{H}{1} cavity. This implies that the \ion{H}{1} cavity protrudes outward to the rear side of the cavity at the southeast region. The detection of the \ion{C}{4} emission line in the southeast region seems to indicate that \object{RCW 114} have evolved relatively slow in this protruded region compared to other regions and the \ion{C}{4}-emitting gas still remains in the region.

\section{CONCLUSIONS}

We have presented the results of the \ion{C}{4} $\lambda\lambda$1548, 1551 emission line detection for the region of \object{RCW 114} using the data set of FIMS/SPEAR. We revealed an enhanced \ion{C}{4} region, which is closely associated with one of the H$\alpha$ filamentary features of \object{RCW 114}. We argued that \object{RCW 114} is not associated with \object{WR 90} and is likely to be much closer to us than the star, based on the \ion{C}{4} luminosity estimation. We also found from the SGPS \ion{H}{1} 21 cm analysis that there is another smaller \ion{H}{1} bubble centered on \object{WR 90} that is not related to the larger \ion{H}{1} bubble associated with \object{RCW 114}. The enhanced \ion{C}{4} region is located near the sharp H$\alpha$ filament in the southeast while no enhanced \ion{C}{4} feature is seen near the broad H$\alpha$ filament in the west. Also, the \ion{H}{1} cavity pre-existing \object{RCW 114} appears to protrude outward to the rear side of the cavity at the southeast region, where it is well overlapped with the enhanced \ion{C}{4} region. These morphologies can be well explained by the scenario in which \object{RCW 114} is an old SNR which has evolved in a non-uniform ambient ISM with a lower-density southeast region and happens to be located along the same line of sight toward the \ion{H}{1} bubble around \object{WR 90}.

\acknowledgments

FIMS/SPEAR is a joint project of the Korea Advanced Institute of Science and Technology, the Korea Astronomy and Space Science Institute, and the University of California at Berkeley, funded by the Korean Ministry of Science and Technology and the National Aeronautics and Space Administration (NASA) Grant NAG5-5355. We thank Naomi Melissa McClure-Griffiths for providing the SGPS data.

\end{document}